\RequirePackage{lineno}
\documentclass[10pt,aps,prb,citeautoscript,preprint,preprintnumbers,amsmath,amssymb,longbibliography]{revtex4-1}

\usepackage{graphicx}
\usepackage[bookmarks=false]{hyperref}
\usepackage{amsmath}
\usepackage{amssymb}
\usepackage{bm}
\usepackage{amsthm}
\usepackage{amsfonts}
\usepackage{latexsym}
\usepackage{wrapfig}
\usepackage[usenames,dvipsnames]{color}
\usepackage[protrusion=true,expansion=true,final]{microtype}
\usepackage{ifthen}
\usepackage{tikz}\usetikzlibrary{shapes,arrows,positioning}
\usepackage{color,soul}
\usepackage{lineno}

\usepackage{chemformula}

\usepackage{caption}

\begin{document}

\title{The shift photovoltaic current and magnetically-induced bulk photocurrent in piezoelectric sillenite crystals}
\author{Aaron M. Burger$^{1}$}
\author{Lingyuan Gao$^2$}
\author{Radhe Agarwal$^3$}
\author{Alexey Aprelev$^4$}
\author{Jonathan E. Spanier$^{1,3,4,5}$}
\author{Andrew M. Rappe$^{2}$}
\email[ ]{email: rappe@sas.upenn.edu}
\author{Vladimir M Fridkin$^{1,6}$}

\affiliation{$^1$Department of Electrical \& Computer   Engineering,\!
	Drexel University,\! Philadelphia,\! PA 19104-2875,\! USA}%
\affiliation{$^2$Department of Chemistry,\!
	University of Pennsylvania,\! Philadelphia,\! PA 19104-6323,\! USA}%
\affiliation{$^3$Department of Materials Science \& Engineering,\!
	Drexel University,\! Philadelphia,\! PA 19104-2875,\! USA}%
\affiliation{$^4$Department of Physics,\!
	Drexel University,\! Philadelphia,\! PA 19104-2875,\! USA}%
\affiliation{$^5$Department of Mechanical Engineering \& Mechanics,\!
	Drexel University,\! Philadelphia,\! PA 19104-6323,\! USA}%
\affiliation{$^6$Shubnikov Institute for Crystallography,\!
	Russian Academy of Sciences,\! Moscow,\! Russian Federation}%
\date{\today}
\begin{abstract}
Recently, it has been shown how shift and ballistic currents in piezoelectric sillenite crystals \ch{Bi_{12}GeO_{20}} and \ch{Bi_{12}SiO_{20}} can be separated experimentally under the assumption that the shift component of the circular current is small. However, it has been claimed that the shift and ballistic currents cannot be quantified by this method, due to the magneto-photovoltaic effect caused either by the change of the crystal’s spatial symmetry in the magnetic field or by the breaking of time-reversal symmetry. Presently, we report observations of photovoltaic currents in \ch{Bi_{12}SiO_{20}}, excited by linearly- and circularly-polarized light under weak external magnetic field, as well as measurements of the corresponding photo-Hall signals. We demonstrate that the magneto-photovoltaic current constitutes a significant fraction of the measured current in the Hall direction for \ch{Bi_{12}SiO_{20}} under specific experimental conditions. 
\end{abstract}

\maketitle
%
The bulk photovoltaic effect (BPE) in non-centrosymmetric crystals has been investigated in many ferro- and piezoelectric crystals.\cite{Fridkin_2013,Sturman_1992} The BPE is a widely-explored phenomenon in non-centrosymmetric materials because it violates the principle of detailed balance and because its application in nanoscale electrode geometries has yielded unity or higher quantum efficiency, even surpassing the Shockley-Queisser limit of power conversion efficiency.\cite{Spanier_2016} The BPE current consists generally of two parts: ballistic $j_b$\cite{Belinicher_1978,Gu_2017} and shift current $j_{sh}$.\cite{Belinicher_1982} They flow in the same crystallographic directions determined by the third-rank photovoltaic tensor, but can have different signs and relative magnitudes. 

The mechanisms of ballistic and shift current are completely different. Ballistic current $j_b$ is caused by the asymmetry of photo-excitation and recombination of non-thermalized (hot) carriers in non-centrosymmetric crystals, which leads in turn to the asymmetric distribution of carrier momentum in the band. Without magnetic field, asymmetric electron-phonon scattering or electron-impurity scattering leads to asymmetry in the diagonal elements of the density matrix ($\rho_{nnk} \neq \rho_{nn-k}$) as well as deviation from the Fermi-Dirac distribution. Thus, ballistic current is a transport type of photocurrent, which is connected with the Boltzmann kinetics principle. The relaxation time of ballistic current is typically on the order of the phonon lifetime $\tau \approx 1 - 10^{-2} \text{ ps} \geq \Gamma^{-1}$, where $\Gamma=eB/m^*c$ is the Larmor frequency, $m^{\star}$ is the electron mass, and $B$ is the magnetic field, permitting measurement of the Hall effect under illumination by linearly- and circularly-polarized light to obtain the mobility of non-thermalized carriers.\cite{Fridkin_2013} 

The first density functional theory calculations of shift current were reported in Refs.~\onlinecite{young2012first} and \onlinecite{Young_2012}. The shift current $j_{sh}$ is related to the off-diagonal elements in the density matrix. The absorption of a photon and the corresponding change of electron energy is accompanied by its shift in real space of the crystal, and therefore shift current, unlike ballistic current, does not involve carrier transport. Since shift current is due to coherent quantum wave-packet evolution during the absorption process, its relaxation time $\tau$ is on the atomic time scale (sub-fs to fs), preventing observation of the Hall effect because $\tau \ll \Gamma^{-1}$ for any applied magnetic field $B$. The kinetic theory of shift current developed in Ref.~\onlinecite{Belinicher_1982} and later within the framework of density functional theory, has recently attracted renewed attention.\cite{young2012first,Young_2012, morimoto2016topological,cook2017design,rangel2017large,fregoso2017quantitative,chan2019exciton}

The situation becomes more complicated in the presence of a magnetic field; magnetically-induced asymmetric scattering leads to an asymmetric photoelectron generation rate, and this contributes to another type of photocurrent, conventionally referred to as $j_H$\cite{Ivchenko_1984}. Photocurrent $j_H$ is connected with time-reversal symmetry breaking due to the magnetic field. Analogous to $j_b$, $j_H$ is related with the classical motion of carriers and thus can also be classified as ``ballistic current''. However, different from $j_b$ which does not depend on the direction of magnetic field $B$, $j_H$ is only perpendicular to $B$. Within a two-band model, at the weak field limit, it can be approximated that $j_H \approx \mu_{nth}B(x j_{sh})$\cite{Ivchenko_1984}, despite the fact that their physical origins and characteristics are quite different. $\mu_{nth}$ denotes the mobility of the scattered current and is assumed to be the same regardless of the light polarization, and $x$ is a theoretically predicted model-dependent parameter for two simple isotropic bands, depending on the corresponding mobilities and carrier effective masses~\cite{Ivchenko_1984}. The linear dependence of $j_H$ on $B$ makes $j_H$ mixed with the Hall signal of $j_b$, and the sum of these two is the as-measured Hall signal in experiment. Other magneto-photovoltaic effects are possible for magnetically-responsive materials such as paramagnetic or magnetostrictive materials, but the presently-considered mechanism of $j_H$ photocarrier scattering by the magnetic field is perhaps the most general.

It has been presumed that the photovoltaic current excited by circularly-polarized light, circular bulk photovoltaic (BP) current, is pure ballistic\cite{Fridkin_2013}. However, the BP current excited by linearly-polarized light (linear photovoltaic current $j_{\ell}$) is the sum of ballistic and shift components. The existence of shift current follows experimentally from the fact that the mobility measured for linear photovoltaic current, $\mu_{\ell}=j_{\ell}^B/j_{\ell}B$, does not equal the mobility of circular current, $\mu_{c}=j_{c}^B/j_{c}B$ ($\mu_{\ell} \neq \mu_c$). Under linearly-polarized light, $j_{\ell}=j_{\ell b}+j_{sh}$ is the total BP current, and $j_{\ell}^B$  is the sum of the photo-Hall signal of $j_{\ell b}$ and $j_H$. Using the relation between $j_H$ and $j_{sh}$ according to Ref.~\onlinecite{Ivchenko_1984}, $j_{\ell}^B$ has the form \begin{equation} j_{\ell}^B \approx \mu_{nth}B(j_b+x j_{sh}). \end{equation} Under circularly-polarized light, $j_c=j_{cb}$ is the pure ballistic current (phonon-induced), and now the equality $\mu_c = \mu_{nth}$ is permitted by $j_H=j_{sh}=0$. $j_c^B=\mu_cBj_{cb}$ only has the photo-Hall signal of $j_{cb}$. Therefore, the difference between $\mu_c =\mu_{nth}= j_c^B/j_{cb}B$ and $\mu_{\ell}=j_{\ell}^B/(j_{lb}+j_{sh})B$ can help distinguish the ballistic current from shift current.

The first separation of ballistic and shift currents in this manner was performed only recently \cite{Burger_2019} by means of linear and circular BP currents under magnetic field in sillenite cubic piezoelectric crystals Bi$_{12}$GeO$_{20}$ (BGO) and Bi$_{12}$SiO$_{20}$ (BSO). Consistent with the aforementioned assumption, in Ref.~\onlinecite{Burger_2019} it was emphasized that this method of shift and ballistic current separation is valid only in cases when the circular photovoltaic current is mainly or completely ballistic. In addition, the separation of $j_b$ and $j_{sh}$ was previously performed for $x=0$, corresponding to the absence of the magnetically-induced photocurrent  $j_H$\cite{Burger_2019}. This can be considered as the first experimental reveal of shift current existence. If $j_H$ is also present, the uncertainty in parameter $x$ does not permit in the common case determination of the values of $j_b$ and $j_{sh}$. However, the uncertainty of $x$ can be bounded in a simple manner, as the value of $j_H$, changing the Hall signal of non-thermalized carriers, cannot be large.  We note that this parameter $x$ was never observed experimentally.

If the Lorentz Hall signal of the pure ballistic current is denoted by $H_1=\mu_cBj_b$ and the magnetically-induced photocurrent by $H_2=\mu_c B x j_{sh}$, it is reasonable to accept the limit: \begin{equation}
    |H_2/H_1|\leq 1,
\end{equation} 
considering that the effect of magnetic scattering is smaller than that of electron-phonon scattering under a weak magnetic field.  To formulate $j_b/j_{sh}$ we have two conditions: (1) The linear (ballistic and shift) photocurrent is not changed with the magnetic field $B$, so we can always write: \begin{equation}
j_b+j_{sh}=j_{\ell}.
\end{equation}
(2) If we introduce the parameter $r \equiv \mu_{\ell}/\mu_{c}$, according to the definition of $j_{\ell}^B$  and $j_{\ell}$, we obtain the equation: \begin{equation}
    j_b + x j_{sh} \approx r j_{\ell}.
\end{equation}
It was shown experimentally\cite{Burger_2019} that for BGO and BSO $r < 1$. In the general case, the solution of (3) and (4) is impossible due to uncertainty in the value of $x$. However, for a small value of $r$, the separation of $j_b$ and $j_{sh}$ could be accomplished. The existence of shift current follows from $r < 1$, as was experimentally established in Ref.~\onlinecite{Burger_2019}. For a few spectral points, we observed $r<1$ and $r\ll 1$. For these points, we performed our analysis.

From (3) and (4) follows the relation: \begin{equation}
    j_{\ell b}/j_{sh} \approx \frac{x-r}{r-1}
\end{equation}
The modulus of the ratio between the magnetically-induced photocurrent $j_H$ and phonon-induced ballistic current $j_{\ell b}$ in accordance with (5) is \begin{equation}
|H_2/H_1|=|x j_{sh}/j_{\ell b}|\approx \left|x \frac{(r-1)}{(x-r)}\right |.
\end{equation}
Thus, a set of measurements made at low magnetic fields can be used to deduce the strength of magnetically-induced photocurrent under the conditions and assumptions outlined above.
%

\begin{figure}
    \centering
    \includegraphics{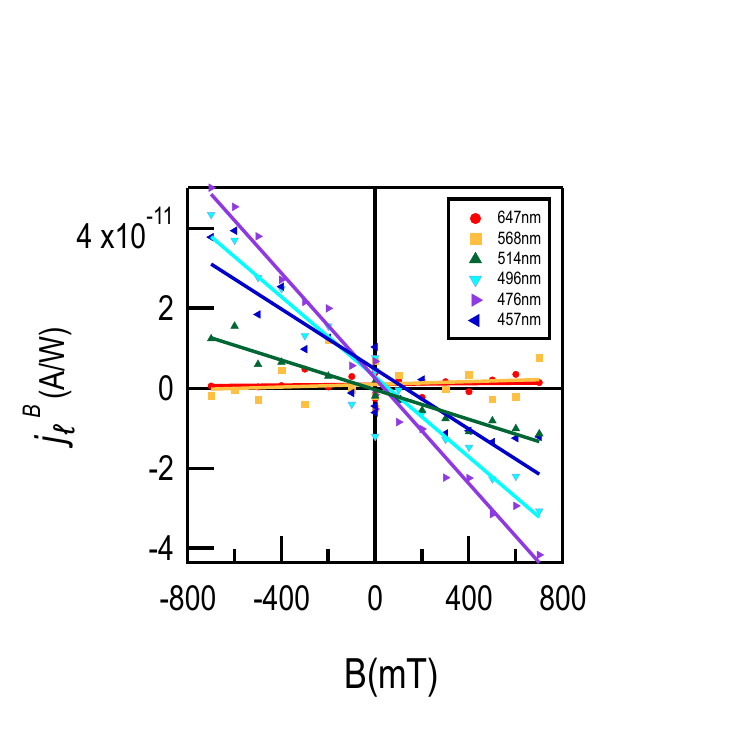}
    \includegraphics{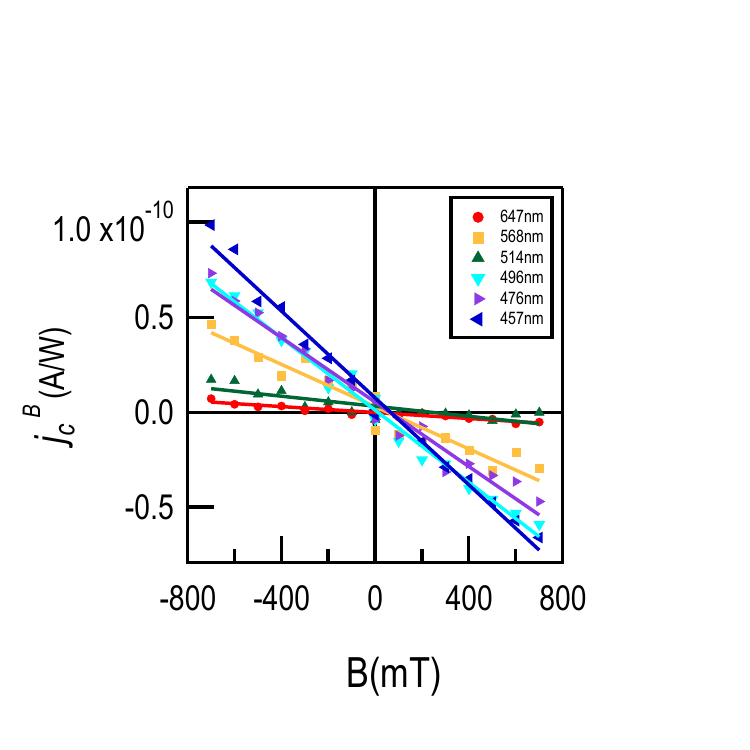}
    \caption{Magnetic field dependence of \textbf{a)} as-measured linear Hall signal at selected wavelengths, and \textbf{b)} as-measured circular Hall signal at selected wavelengths. Currents are normalized to the incident power.}
    \label{fig:my_label0}
\end{figure}

\begin{figure}
    \centering
    \includegraphics{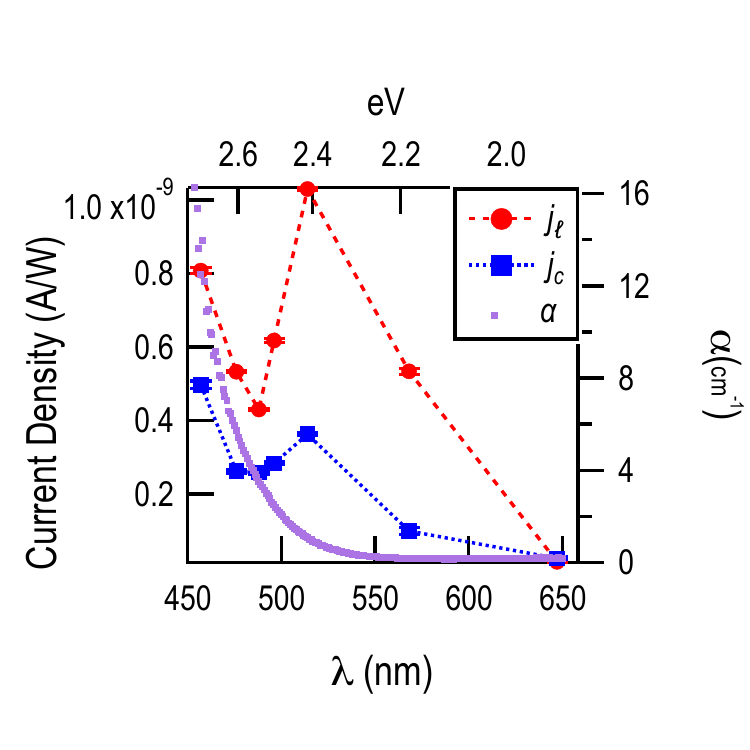}
    \includegraphics{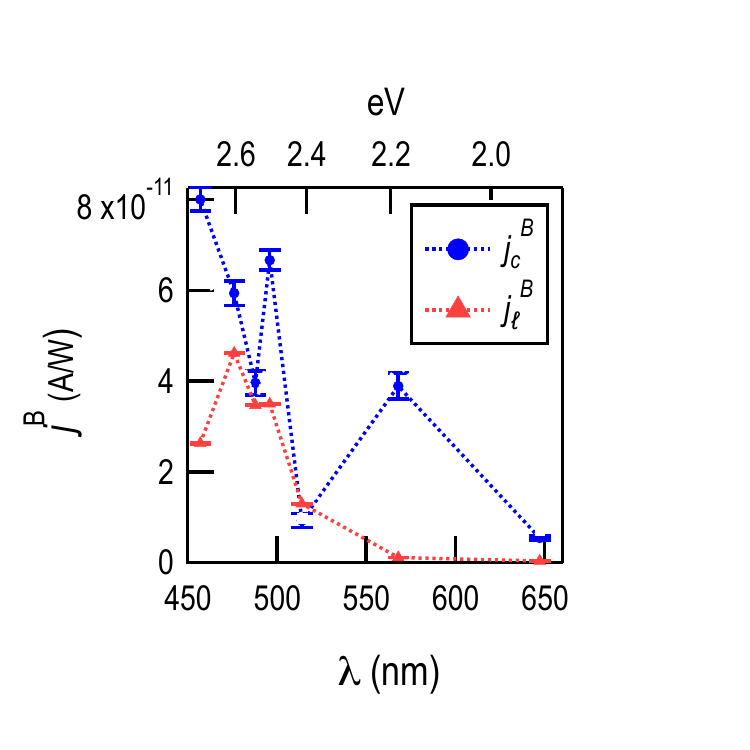}
    \caption{Dependence of \textbf{a)} linear and circular photovoltaic current, and absorption coefficient $\alpha$, and \textbf{b)} linear and circular photo-Hall currents at selected wavelengths. Currents are normalized to incident power.}
    \label{fig:my_label}
\end{figure}

\begin{figure}
    \centering
    \includegraphics{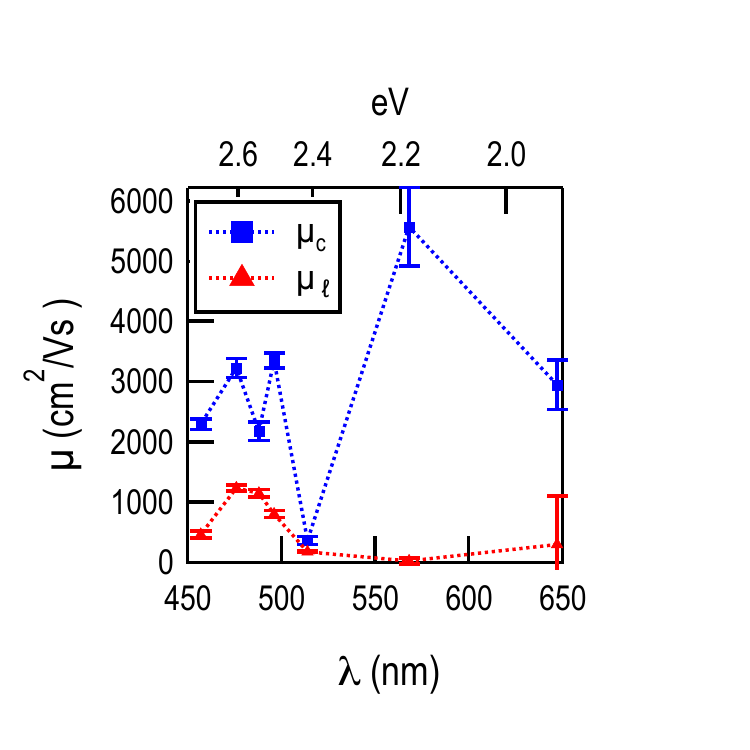}
    \includegraphics{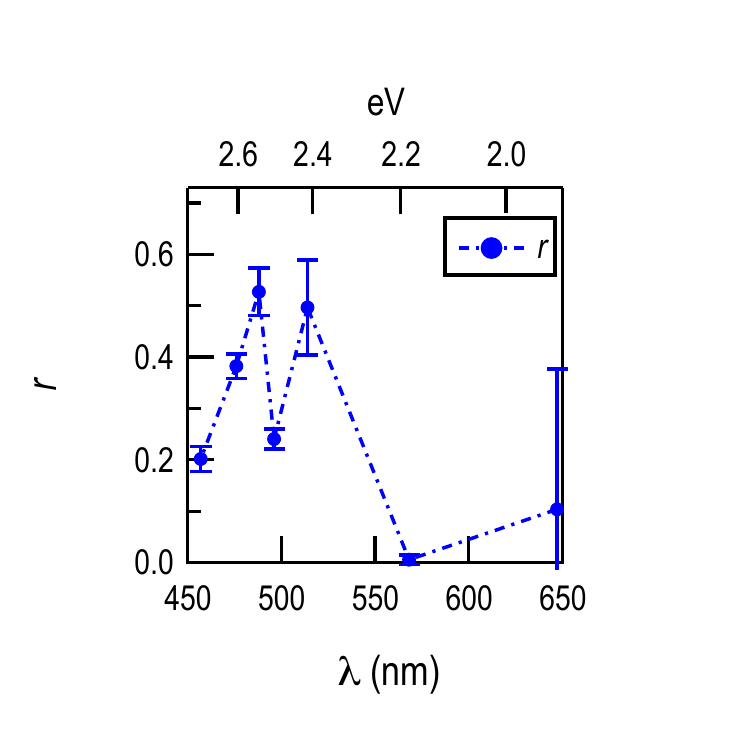}
    \caption{\textbf{a} Non-thermalized mobilities $\mu_{\ell}$ and $\mu_{c}$ for linearly- and circularly-polarized light, respectively, and \textbf{b}, parameter $r\equiv \mu_{\ell}/\mu_c$, at selected wavelengths.}
    \label{fig:my_label2}
\end{figure}

We measured values of photovoltaic and photo-Hall currents under linearly- and circularly-polarized light in single-crystal gyrotropic cubic piezoelectric \ch{Bi_{12}SiO_{20}} (symmetry point group T) having one linear component $G^{\ell}_{14}$ and one circular component $G^c_{11}$ of the bulk photovoltaic tensors $G^{\ell}_{ikl}$ and $G^c_{il}$\cite{Sturman_1992}, respectively. Measurements were performed at room temperature under polarized monochromatic Ar-Kr ion laser irradiation of intensity $\approx$ 132-777 mW cm$^{-2}$, using commercial single crystals (MetaLaser Photonics, Nanjing CHN). The circular and linear currents were extracted by electro-optic modulation using a Pockels cell (Conoptics, Danbury CT). The measurements of Hall components of linear and circular current were performed under magnetic field $B$ = 0.7 T (Montana Instruments, Bozeman MT). The energy gap of the sillenite is $\approx$ 3.2 eV, and we performed measurements in the extrinsic impurity region of 450-650 nm. The photovoltaic $j_l$ and $j_c$ were obtained through radio-frequency sputtered indium tin oxide transparent electrodes along one of the $<$100$>$ directions. The method of measurements was described in detail elsewhere \cite{Burger_2019}. As follows from above, the measured photovoltaic current has extrinsic character and is caused by the excitation of electrons from impurities to the conduction band. The photoconductivity in BSO (like BGO) is $n$-type, arising from Bi$^{+3}$ donors, and the mobility of thermalized electrons $\mu_{th}$ is very small (10$^{-2}$-10$^{-6}$ cm$^2$ V$^{-1}$ s$^{-1})$\cite{Kostyuk_1980,Frejlich_2007}. Fig.~\ref{fig:my_label0} shows the magnetic field dependence of the as-measured linear and circular Hall currents. The linear dependence of the photo-Hall signal on the magnetic field $B$ is demonstrated clearly for both scenarios, and the slopes are related with non-thermalized mobilities and corresponding photocurrents. In Fig.~\ref{fig:my_label} we show the measured linear and circular photovoltaic currents at selected wavelengths and the optical absorption in BSO, along with the measured linear and circular photovoltaic Hall currents. From photovoltaic and Hall currents, the non-thermalized mobilities $\mu_{\ell}$ and $\mu_{c}$ can be calculated, and they exceed $\mu_{th}$ by a few orders of magnitude (Fig.~\ref{fig:my_label2}a), yielding the parameter $r=\mu_{\ell}/\mu_{c}$, which is seen to vary for different wavelengths measured (Fig.~\ref{fig:my_label2}b). From Fig.~\ref{fig:my_label2}b it is seen that at least for three spectral points $r\ll1$ or $r \approx 0.2$, and our following discussions are based on these points.

%

Let us consider $r$ in (4) as a small parameter $(0<r \ll 1)$. Here, we discuss three main cases:
\begin{enumerate}
    \item $|x|\gg r$, $r \ll 1$. Following (3), (4) and (5), $j_{\ell b}/j_{sh}\approx -x$, $j_{\ell b}$ is cancelled with $j_{sh}$ when $x \approx 1$. $j_{\ell}=(1-x)j_{sh}$ is the linear photovoltaic current. The ratio of Hall signals is $|H_2/H_1|\approx 1$. This is attainable when the magnetic scattering is at the same order as the phonon scattering.
    \item $|x|\ll r$. In this case $j_b/j_{sh}\approx r$, $j_{sh} \gg j_b$ and $|H_2/H_1|\ll 1$. This is a very unlikely case, especially for an impurity band transition, since the asymmetric electron-impurity scattering should induce a large $j_{\ell b}$.
    \item $|x|$ and $r$ are of the same order. In this case $j_b/j_{sh}\approx r-x$. $j_{sh}\gg j_b$, $|H_2/H_1| \gg 1$. The directions of $j_{sh}$ and $j_b$ can be opposite or the same depending on the sign of $r-x$. In this case, the BP current is dominated by $j_{sh}$, and the effect of asymmetric scattering is very small. Similar to case 2, this is also very unlikely for an impurity band transition.
\end{enumerate}

In conclusion, we emphasize once more that the existence of shift photovoltaic current directly follows from $\mu_{\ell} \neq \mu_c$. As seen from (5), in the general case the experimental parameter $r = \mu_{\ell}/\mu_{c}$ does not permit precise determination of values of $j_{sh}$ and $j_b$ due to the uncertainty of $x$. But the degree of smallness of $r$ in the extrinsic spectral region makes this separation more robust and quantitative. Based on the above discussions of the three cases, it is most likely that $|j_b|\approx x|j_{sh}|$ for an impurity band transition. For this case, we provide theoretical and experimental evidence that the magnetically-induced scattering in BSO is significantly large, of the same order as the phonon scattering. 

\begin{acknowledgments}
Work at Drexel University was supported by the National Science Foundation, Division of Chemical, Bioengineering, Environment and Transport Systems, under grant number CBET 1705440, and the Division of Materials Research, under grant number DMR 1608887. The theoretical work, by L. G. and A. M. R., was supported by the Department of Energy, Office of Science, Office of Basic Energy Sciences, under grant number DE-FG02-07ER46431.
\end{acknowledgments}



%

  
\end{document}